\documentstyle[preprint,aps,psfig]{revtex}
\begin{document}
\draft

%%%%%%%%%%%%%%%%%%
%\twocolumn[\hsize\textwidth\columnwidth\hsize\csname
%@twocolumnfalse\endcsname
%%%%%%%%%%%%%%%%%%

\widetext
\title{  Green Function Monte Carlo  with Stochastic Reconfiguration  }
\author {  Sandro Sorella }
\address{ 
 Istituto Nazionale di Fisica della Materia and
  International School for Advanced Studies 
Via Beirut 4, 34013 Trieste, Italy 
  } 
\date{\today}
%\date{\today, SISSA preprint:145/97/CM/MB}
\maketitle
\begin{abstract}
 A new method for the stabilization of the sign problem in   the  Green Function
Monte Carlo  technique is proposed. The method is devised for real lattice
Hamiltonians and is based on an iterative ''stochastic reconfiguration'' scheme
which  introduces  some bias   but allows a stable simulation with constant sign.
The systematic reduction  of this bias is in principle possible. 
The  method is applied to the frustrated $J_1-J_2$ Heisenberg model, and tested
against  exact diagonalization data.  Evidence of a finite   spin gap for
$J_2/J_1 >\sim 0.4$  is found in the thermodynamic limit.  
\end{abstract}

%%%%%%%%%%%%%%%%%%
%]
%%%%%%%%%%%%%%%%%%

\narrowtext

As well known the  Green Function 
 Monte Carlo method (GFMC)  allows to obtain  the  exact 
ground state properties of a many body hamiltonian with a statistical 
method. 
 One of the most severe restriction is
that only  positive definite Green function GF  can be sampled, otherwise 
the method is facing the well known ''sign problem''.
Approximate techniques like the fixed node approximation  (FN) 
 have  been developed to circumvent the sign  problem 
but at the very least  
they cannot be systematically improved to achieve the exact 
answer within statistical errors. 
This  property has  severely   limited  the  applications  
of GFMC to fermions and frustrated boson models.
In this letter I propose  a  new  approach to stabilize  
the sign problem, the   
 GFMC  with stochastic reconfiguration  (GFMCSR),
which  will be shortly described below, revisiting also the basic steps of 
the standard GFMC on a lattice.\cite{ceperley,runge}

In order to filter out  the ground state of
a given lattice hamiltonian $H$  
the standard power method may be  applied iteratively :
\begin{equation} \label{iter} 
\psi_{n+1} (x^\prime) = 
\sum_x (\Lambda \delta_{x^\prime,x} -  H_{x^\prime,x})  \psi_n(x)
\end{equation}
where $x$ represents conventionally the index of a complete basis $|x>$ 
, $H_{x^\prime,x}$ being the corresponding 
matrix elements of the hamiltonian which  in the following are assumed real, 
and $\Lambda$ is a positive constant   that allows the   convergence of $\psi_n$ 
to the ground state $\psi_0(x)$,  for large $n$.
In numerical calculations of interesting lattice hamiltonians
the dimension of the basis  grows exponentially   
with the size and the particle number, 
though the matrix itself is very sparse  and   all
its   elements $H_{x^\prime,x}$ , for given $x$, can be generally computed
even for  large  system size.
In this case   an exact application of  (\ref{iter}) is
impossible  unless for few steps.
 A way out is to use a  stochastic approach , like GFMC 
,which   is particularly simple on a lattice. 
 
 In order  to implement stochastically the iteration (\ref{iter}) the 
corresponding lattice GF 
\begin{equation}\label{green}  
G_{x^\prime,x}=\Lambda  \delta_{x^\prime,x} -  H_{x^\prime,x} 
\end{equation} 
may be decomposed  in  the following way:
\begin{equation} \label{defgfmc}
 G_{x^\prime,x}= s_{x^\prime,x} p_{x^\prime,x}  b_x 
\end{equation}
where $p_{x^\prime,x}$ is  a  normalized stochastic matrix,  
$b_x \ge 0$ is a normalization 
constant and the matrix $s$ takes into account the sign of the GF.
The typical  choice is to take $ p_{x^\prime,x}= |G_{x^\prime,x}|/b_x$ , 
$b_x= \sum_{x^\prime} |G_{x^\prime,x}| $ and 
$s_{x^\prime,x}= {\rm sgn}\, 
  G_{x^\prime,x} $, which is identically one if there is 
no sign problem.

 In the GFMC  method the so called  ''walker``  
 is defined by a weight $w$ and  a configuration $x$.. 
At a given  iteration $n$ the walker is assumed to sample statistically 
the state $\psi_n(x)$ in Eq.(\ref{iter}), in the sense that  the 
probability $P_n(w,x)$ to have the walker with weight $w$ (not restricted to be 
positive) in a given  configuration $x$ satisfies:
$\int  dw P_n(w,x)  w = \psi_n(x).$
Then the matrix multiplication (\ref{iter}) can be implemented statistically 
, in the precise sense that 
$\int   dw P_{n+1} (w,x)  w =\psi_{n+1} (x)$,
by the following three steps:
\begin{enumerate} 
\item scale the walker weight  by  $b_x$: $ w^\prime  = b_x w $.
\item select    randomly  a new configuration $x^\prime$ according to the
stochastic matrix $p_{\displaystyle x^\prime,x}$.
\item finally multiply the weight of the walker by the sign factor
 $s_{\displaystyle x^\prime,x}$:
$ w^\prime \to w^\prime s_{\displaystyle x^\prime,x}$ ~~~~~~~   (MI)
\end{enumerate}
In principle the previous Markov process determines, for large 
$n$, the ground state of  $H$ even with a single walker. 
In practise it is  convenient to use a large number $M$ of 
 walkers, which I indicate by  
$(w_j, x_j)$ $j=1,\cdots M$,  shorthand in the following also by vector
notations  $\underline{w},\underline{x}$.

  If there is sign problem the average walker sign 
$<s>_n ={ < \sum_j w_j >_n  \over <  \sum_j |w_j|>_n   } $
 decreases exponentially to zero as the Markov iteration MI  is
repeatedly applied and it is basically impossible  to reach a
reasonably large value of $n$.
%Henceforth the brackets $< >_n$ indicate the 
%average over the distribution $P_n$.
 
 Recently  a remarkable progress in GFMC on a lattice was the extension 
of the FN to this case. 
The method is based on a definition of an effective GF  
$G^{f}_{x^\prime,x}$ 
which is always positive definite  but yields a good {\em variational}
estimate of the energy.   For later purposes  
we  define this effective GF 
in a slightly different way, by introducing a parameter $\gamma$:
which allows to sample also the negative elements of the GF : 
\begin{equation} \label{geff} 
G^{f}_{x^\prime,x}= \left\{
\begin{array}{ccc}
 - H_{\displaystyle x^\prime,x} &
 {\rm if} & H_{\displaystyle x^\prime,x}\le 0  \\
\gamma H_{\displaystyle x^\prime,x}
 & {\rm if} & H_{\displaystyle x^\prime,x}>0 \\
 \Lambda -H_{x,x} - (1+\gamma) {\cal V}_{\rm sf}  (x)  & {\rm if}& 
 x=x^\prime  
\end{array} \right.
\end{equation}
where  the diagonal {\sl sign-flip} contribution  
is given by\cite{bemmel,ceperley1}:
\begin{equation} \label{signflip}
 {\cal V}_{\rm sf} (x) =   \sum\limits_{H_{\displaystyle x^\prime, x} > 0, 
x^\prime \ne x }
H_{\displaystyle x^\prime,x}
\end{equation}
For $\gamma=0$ the usual formulation\cite{ceperley1} is recovered, whereas for 
$\gamma>0$\cite{theorem}  the crossing to the negative sign region is 
allowed so that the exact GF 
 can be written as $G_{x^\prime,x}=
  s_{x^\prime,x} G^{f}_{x^\prime,x}$ where  $s_{x^\prime,x}$ 
is finite and non zero and is determined  by the ratio 
$G_{x^\prime,x} \over G^{f}_{x^\prime,x} $ with $G$ and $G^{f}$ given 
by Eq.(\ref{green}) and Eq.(\ref{geff}) respectively.
The value of the constant $\gamma$ necessary to cross the ''nodal surface''
was chosen to be $1/2$ in all forthcoming applications.

%\begin{equation} \label{sxprimex}
%s_{\displaystyle x^\prime,x}= \left\{
%\begin{array}{ccl}
%1 &  {\rm if} & H_{\displaystyle x^\prime,x}\le 0 \\
%-1/\gamma & {\rm if} & H_{\displaystyle x^\prime,x}>0 \\
%{ \Lambda - H_{\displaystyle x,x } \over
% \Lambda -   H_{\displaystyle x,x } - (1+\gamma)   {\cal V}_{\rm sf} (x) }
%& {\rm if} & x^\prime=x
%\end{array} \right.
%\end{equation}
In the basic decomposition (\ref{defgfmc})  the stochastic matrix
 $p_{x^\prime,x}=G^{f}_{x^\prime,x}/b_x$
and the  normalization coefficient $b_x=\sum_{\displaystyle x^\prime}
G^{f}_{x^\prime,x}$ are instead determined  only by $G^{f}$.  

 By omitting  the last step $w^\prime \to w s_{x^\prime,x}$ in
  the Markov iteration process MI, the state  $\psi_n$ is indeed propagated
 trough the
positive  GF $G^{f}$.
 The main property  used in the following  is that at any
Markov iteration $n$ we can have a statistic  knowledge   of  both the state 
$\psi_n(x)$ obtained  with  the
exact  GF and of  $\psi_n^{f} (x)$ obtained instead with  the
approximate but  positive definite one $G^{f}$. 
To this purpose  the $j^{th}$ walker is defined by  
two  weights $w^{f}_j$ and
$w_j$ corresponding  to the propagation of the walker by 
$G^{f}$ and $G$ respectively.
 These weights 
act on {\em the same } configuration $x_j$. 
Hereafter the vector $\underline{w}$ represents therefore a shorthand 
notation for   the $2 M$
 components $w_j,w^f_j$ for  $j=1,\dots M$.

The walker vector  
$\underline{w},\underline{x}$ 
allows to determine statistically the state:
\begin{equation} \label{state}  
   \psi_n(x ) =
 \int d [\underline{w}] \, \sum_{\underline{x}} \,
P_n(\underline{w},\underline{x})  
 \,\,  \sum_j  \delta_{x,x_j}\, w_j/{\scriptstyle M}    
 \end{equation} 
and analogously $\psi^f_n(x)$ by replacing the weights $w_j$ with 
the positive ones $w^f_j$ in the previous equation. 
In this way  the configurations generated by the described Markov 
process MI, if   weighted with the constants   $w^f_j$,  are distributed 
for large $n$,
according to the variational  state corresponding to  $G^{f}$. This is  
 a reasonable variational wavefunction (WF), which 
will be the initial approximation to which systematic corrections will be
applied, as described later on. 

%sss first change 
  Apart for the previous technical definitions, we can explain in few 
words  the  basic idea used for the stabilization of the sign problem, 
The iteration MI  converges to the ground state, but due to the sign problem,  
only few iterations can be performed  with a reasonable statistical accuracy.  
However,
%  following the main ideas of  previous  
% stabilization techniques,\cite{koonin,sandro}  
 the representation of the state $\psi_n(x)$ in terms of the walker
  population $x_j,w_j$ is not unique. In fact  it is perfectly  possible to
represent the same state $\psi_n(x)$ either with a walker population  with very
small average  sign  or with
a one with a very large average sign.
\noindent If such  reconfigurations are possible each few $k_p$ steps,
 the average  sign   may be stabilized to a large value during
the  iteration  (\ref{iter}) and there will be no difficulty to  sample the
ground state for $n \to \infty$,  {\em with no sign problem}.

 I will show that this reconfiguration is well defined and indeed possible.   
  The set of  $M$ walkers  
$( \underline{w}, \underline{x})$ are  defined via their probability function 
$P_n(\underline{w},\underline{x})$ which in turn defines the 
state  $\psi_n (x) $ 
 by Eq.(\ref{state}).
 The task is to change $P_n$ onto a new probability distribution $P_n^\prime$ 
corresponding to a steadily high  sign for the walker population. This 
{\em without changing the information  
content}, the state  $\psi_n(x)$.

 Let us define the 
new state $\psi^\prime_n (x)$, as the one obtained by  averaging over 
$P^\prime_n$  in Eq.(\ref{state}), then the reconfiguration 
is exact if $P^\prime_n$ is such that:
\begin{equation} \label{exsrcon}
\psi^\prime_n(x) =\psi_n(x) ~~  {\rm for~ all~}  x
\end{equation}
In general  it is difficult or impractical to realize all these 
conditions (\ref{exsrcon}) as their number equals the dimension 
of the Hilbert space.
  I consider therefore a set of  operators $O^k$, $k=1,\cdots p << M$  and
require only $p+1$    stochastic reconfiguration conditions:
 \begin{equation}
\label{srcon}  \sum_{x^\prime,x}  O^k_{x^\prime,x} \psi^\prime_n(x)= 
\sum_{x^\prime,x}  O^k_{x^\prime,x}  \psi_n (x) 
\end{equation} 
for $k=1,\cdots  p$,  beyond the 
normalization one $\sum_x \psi^\prime (x)=\sum_x  \psi_n(x)$

 The previous equations (\ref{srcon}) mean that the 
so called ''mixed averages'' of the operators $O^k$ coincide before 
and after the reconfiguration.\cite{guiding} 
%the value of the mixed estimate on a
%given  walker  configuration $x_j$ being  given by 
%$O^k_j= \sum_{x^\prime} O^\prime_{x^\prime,x_j}$.

The main  idea of this work is that these $p+1$ conditions can 
be fulfilled {\em exactly}  (for chosen operators)  by defining the 
reconfiguration in the following form:  
\widetext
\begin{equation} \label{reconfiguring}
 P^\prime_n ( \underline{w}^\prime,
 \underline{x^\prime})= \int d [ \underline{w}] \,  
 \sum_{\displaystyle \underline{x}} 
\prod\limits_{\displaystyle i=1}^M \left\{ { \sum_j |p_{x_j}|
\delta_{\displaystyle \displaystyle x^\prime_i,x_j}
 \over \sum_j |p_{x_j}|   } \, \delta ( w^\prime_i -
 { \sum_j w_j \over \beta M}  {\rm sgn}\, p_{x^\prime_i} )
\,\, \delta ( w^{f\,\,\prime}_i - |w^\prime_i|) \right\}
P_n ( \underline{w},\underline{x})
\end{equation}
\narrowtext
\noindent where $\beta={\sum_j p_{x_j} \over  \sum_j |p_{x_j}| }$ 
is the average 
sign  after the reconfiguration
 which is supposed to be much higher to stabilize the process.
The new configurations $x^\prime_i$
are taken randomly among the old ones $\left\{ x_j \right\}$, according to the
table $p_{x_j}$, defined below.. 
  The positive weights  $w^{f}_j$ represent a good starting point  
 for the definition of    a reconfiguration 
with large $\beta$. Though there is some arbitrariness in 
the definition of the coefficients $p_{x_j}$, 
 a simple and convenient choice is:
$$p_{x_j} =w^{f}_j ( 1 +\sum_k \alpha_k ( O^k_j - \bar O^k_{f} ) ) $$
where $ \bar O^k_{f} = { \sum_j w^{f}_j  O^k_j  \over 
\sum_j  w^{f}_j } $ are the  averages over the positive weights  $w^{f}_j$
of the mixed estimates  $O^k_j= \sum_{x^\prime} O^k_{x^\prime,x_j}$  
   corresponding to  the operator $O^k$ and the  configuration $x_j$. 
  
Then, in order to satisfy  the WF conditions (\ref{srcon}), by using the
definition (\ref{reconfiguring}), it is {\em sufficient } that the coefficients
$p_{x_j}$  satisfy the following Markovian conditions:
\begin{equation} \label{condition}
{ \sum\limits_j p_{x_j} O_J^k   \over \sum\limits_j p_{x_j} } =
{ \sum\limits_j  w_j O_j^k \over \sum\limits_j  w_j }
\end{equation}
which in turn determine  the  
unknown variables $\alpha_k$, for $k=1,\cdots p$, for given
$\underline{w},\underline{x}$.

For hamiltonian not affected by the sign problem ($G^{f}=G$ $\alpha_k=0$ and
$\beta=1$)  this  reconfiguration was already used 
 to  control the walker population size without introducing any source of 
systematic error.\cite{calandra}  
 The present  more general reconfiguration 
(\ref{reconfiguring})  can be easily
and efficiently implemented in a similar way.

   Obviously  the reconfiguration conditions (\ref{srcon})  
 are equivalent  to  the {\em exact} 
ones (\ref{exsrcon}),  when the number $p$ of linearly independent operators  
considered in (\ref{srcon}) is equal to  the large   
 dimension of the Hilbert space. 
%sufficiently large but not infinite. 
%The  proof of this important statement is very simple. Consider first the
%diagonal operators. All these operators may be written as linear combinations of
%the ''elementary'' ones 
%$O^{x_0}_{x^\prime,x}
%=\delta_{x^\prime,x} \delta_{x,x_0}$
%acting on  a single  configuration
%$x_0$,  plus at most some constants.  If 
% condition (\ref{srcon}) is satisfied  for  {\em all}  the  elementary
%operators it immediately follows that  $\psi^\prime_n  (x_0)=
% \psi_n (x_0) $ for
%all $x_0$, which is the  exact stochastic reconfiguration
%condition (\ref{exsrcon}). 
%   Then it is simple to show that the coefficients
% $p_{x_j}$, determining $P^\prime$ and $\psi^\prime$, 
% are invariant  for any constant shift   
% of the  operators $O^k$.  Further with  a little algebra it
% turns out that  these coefficients $p_{x_j}$ do not change  for any
% arbitrary   linear transformation of the chosen operator set:  $O^{k \prime}  =
% \sum_{k} L_{k^\prime,k} O^k$ (with real $L$ and $\det L \ne 0$). 
% Thus the proven 
% convergence of the GFMCSR  is   obtained for any sequence of  
% diagonal  operators, that, with increasing $p$, becomes complete. 
%For non-diagonal operators $O_{x^\prime,x}$   note simply that they assume 
%the same mixed average values  of the equivalent diagonal ones
%$O^{f}_{x^\prime,x}=\delta_{x^\prime,x}  \sum_{x^\prime} O_{x^\prime,x}$. 
%Thus the proof   that GFMCSR  converges in principle to the exact solution is
%valid in general even when  non diagonal operators, like the energy.  are
%included  in the conditions (\ref{srcon})  $\Box$. 
An important applicative  issue  is  whether GFMCSR converges, within a
reasonable accuracy, even  with a 
small  number $p$ of meaningful operators $O^k$.

We consider the frustrated $J_1-J_2$ Heisenberg spin 1/2 model on a 
finite square lattice with $L$ sites and periodic boundary conditions
(tilted by 45 degrees for the $L=32$ size only).
%% sss changed above 
  The model  hamiltonian is determined
by  an antiferromagnetic  coupling $J_1>0$ between nearest neighbor spins 
 and a frustrating coupling $J_2>0$ between next neighbor 
ones.\cite{dagotto,hatano,poilblanc}
In all forthcoming examples the stochastic reconfigurations  were applied 
frequently enough to maintain the average  sign before reconfiguration 
 $\sim 0.8$,
 condition that minimize the statistical fluctuations. 
Moreover in each simulation it is important to work with a fairly large 
number of walkers, since in the $M \to \infty$ limit, 
the GFMCSR  results are practically independent of the frequency of
reconfigurations,  as well as the overall constant energy shift $\Lambda$. 

The accuracy  of GFMCSR  for  the ground state   is
displayed in  Tab.\ref{taben} 
, and  compared with other  methods.
The variational 
 WF  (used also for GFMC importance  sampling\cite{guiding}) 
contains a Jastrow like factor 
$$Exp( 
 { \eta \over 2} \sum_{R,R^\prime} v(R-R^\prime) S^z_R S^z_{R^\prime}  )$$ 
to mimic the  interaction between the spins $S^z_R=\pm 1/2$ at sites
$R, R^\prime$,  where $\eta$  is a variational parameter and the two-spin
interaction $v$ can be derived by using the method described in  
 \cite{franjic}, yielding an explicit Fourier transform 
for $v$: 
$$v_q/2=  1 -\sqrt{ { 2 - \sigma ( 1 -\cos q_x \cos q_y) + \cos q_x +
\cos q_y\over  2  -\sigma  ( 1 -\cos q_x \cos q_y)  - \cos q_x - \cos q_y }}$$
 with $\sigma=2 J_2/J_1$. 
 This potential is not defined 
for $J_2/J_1=1/2$, and in such case  I have 
chosen to work with $\sigma=0.8$. 
Restriction to any subspace of total spin projection $S^z_{tot} = \sum_R S^z_R$ 
allows to evaluate the spin gap by performing two simulations for $S^z_{tot}=0$ 
and $S^z_{tot}=1$. Henceforth I will use the the same potential $v$ in both
subspaces, by optimizing $\eta$ for the $S^z_{tot}=0$ energy.  

\begin{figure}
\centerline{\psfig{figure=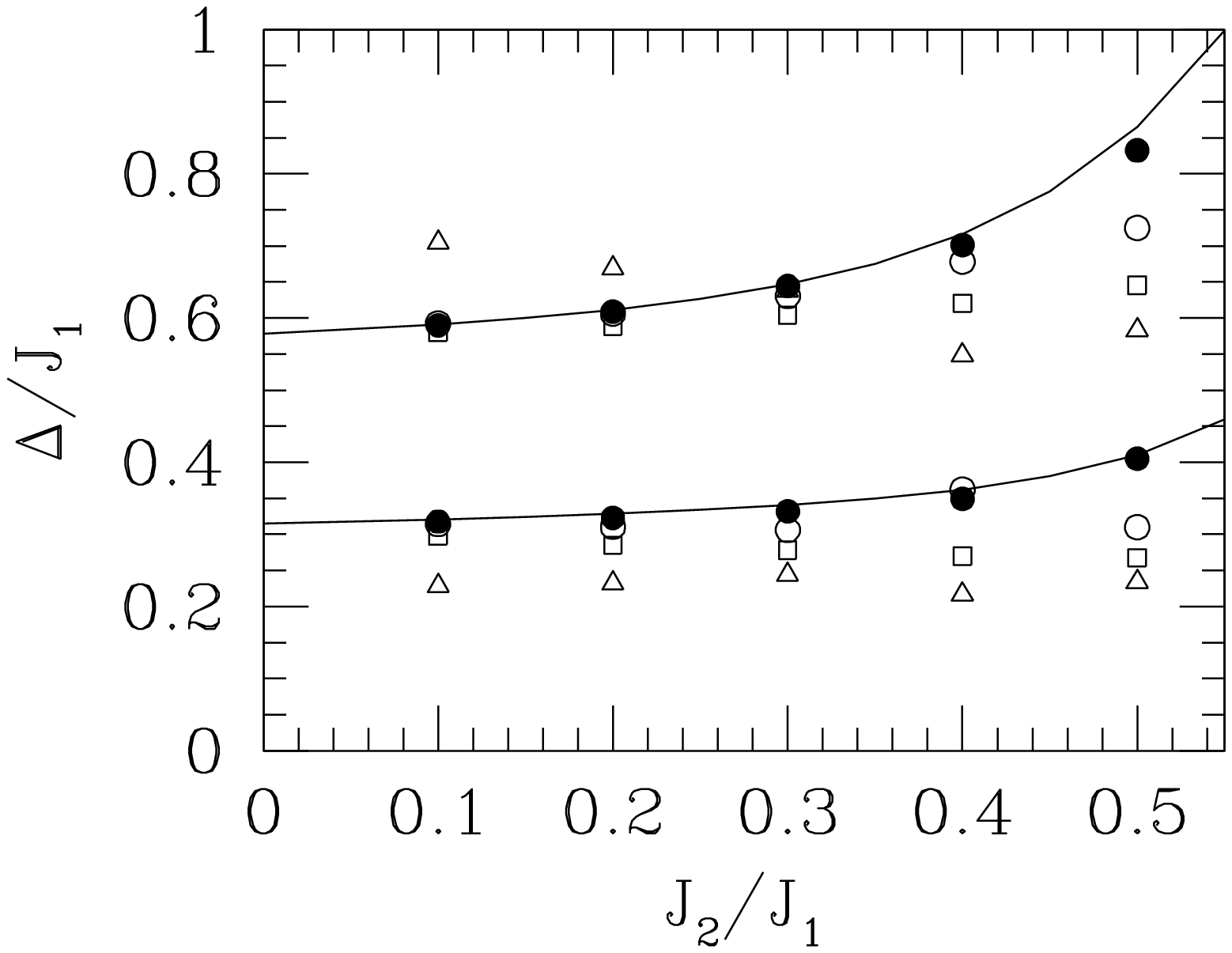,height=13cm}}
\caption {\baselineskip .185in\label{plotgap}
 Estimate of the spin gap for several methods: variational 
(empty triangles), FN  (empty squares), GFMCSR $p=1$ (empty dots)
, GFMCSR (full dots)  as in the table 
for $L=16$ (upper points) and $L=32$ (lower ones).   
 The exact results are connected by continuous lines. 
}
\end{figure}

As shown in the table the accuracy of the variational WF is rather 
poor, and is considerably improved by the FN, at least  for
small  $J_2$. This kind of accuracy is however not enough to determine the 
rapid increase of the spin gap as  $J_2/J_1$  approaches the value 
$1/2$ of  the classical transition. 
 Instead, as shown in  Fig.(\ref{plotgap})
the GFMCSR allows to achieve a good accuracy also on 
this delicate quantity by considering in the reconfigurations  only the energy
and the spin structure factor  $S_q^z=\sum_{R,R^\prime} e^{ i q (R-R^\prime) } 
S^z_R S^z_{R^\prime}$ symmetrized over all directions and for
all non equivalent wavevectors $q$. 
Remarkably also mixed averages  of  correlation functions that are not included
in such  reconfiguration conditions (\ref{srcon}) are also significantly 
improved (see table). 

%The approach of the total spin $\vec S^2$ to its exact integer
%value $S %(S+1)$ also allows to verify the accuracy of the calculation even for
%large size %when exact diagonalization is not possible. The improvement in the 
%%mentioned accuracy is indeed rather independent of size, at least for the 
%%size $L \le 100$ considered here.      

\begin{figure}
\centerline{\psfig{figure=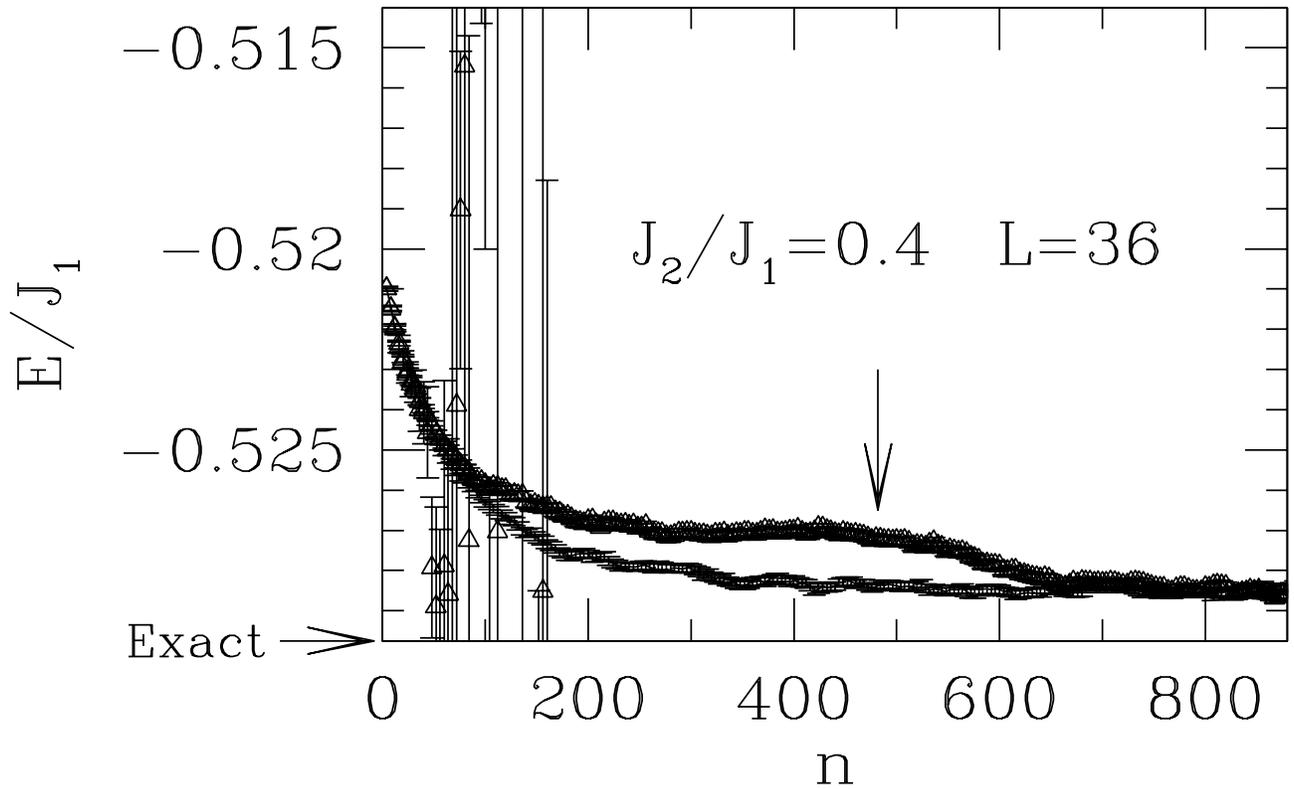,height=11cm}}
\caption {\baselineskip .185in\label{figsign}
 Energy per site vs. n for GFMCSR with $p=1$ (upper  curve to the 
left of the arrow) and 
$p=9$  (remaining  curves). The triangles  represent the 
standard method with  sign problem, i.e. with  large error bars already for
$n>15$. }
\end{figure}

The way GFMCSR reaches the large $n$ limit (at fixed number of operators
$p$) is displayed in Fig.(\ref{figsign})  where 
the initial $n=0$  distribution was obtained by the FN for
$\gamma=0$. For fixed $p$ the algorithm is Markovian and reaches an equilibrium 
distribution for $n\to \infty$, independent of  the initial one (see example 
in Fig.\ref{figsign} where $p$ was changed 
at the iteration indicated by the arrow), 
this in turn will converge to the ground state distribution for 
large $p$.
 A comparison with the standard  ''release nodes''  estimate 
is also shown in the picture. It is clear that there is no hope to   obtain 
meaningful results in this case by the direct sampling of the sign. 
On the contrary this  method looks very stable and, though approximate, a 
convergence to a reasonable accuracy is obtained even with a very small number 
of operators, compared to the dimension of the Hilbert space.

The data shown in the table and in the picture indicate that the 
accuracy  of GFMCSR  may become   rather size independent 
with a relatively small increase of the operator  number $p$. 
The  error to work at finite small $p$ is systematic. 
Thus there is a considerable cancellation of this error for 
 the  determination of the spin gap displayed in  Fig.(\ref{plotgap}).

\begin{figure}
\centerline{\psfig{figure=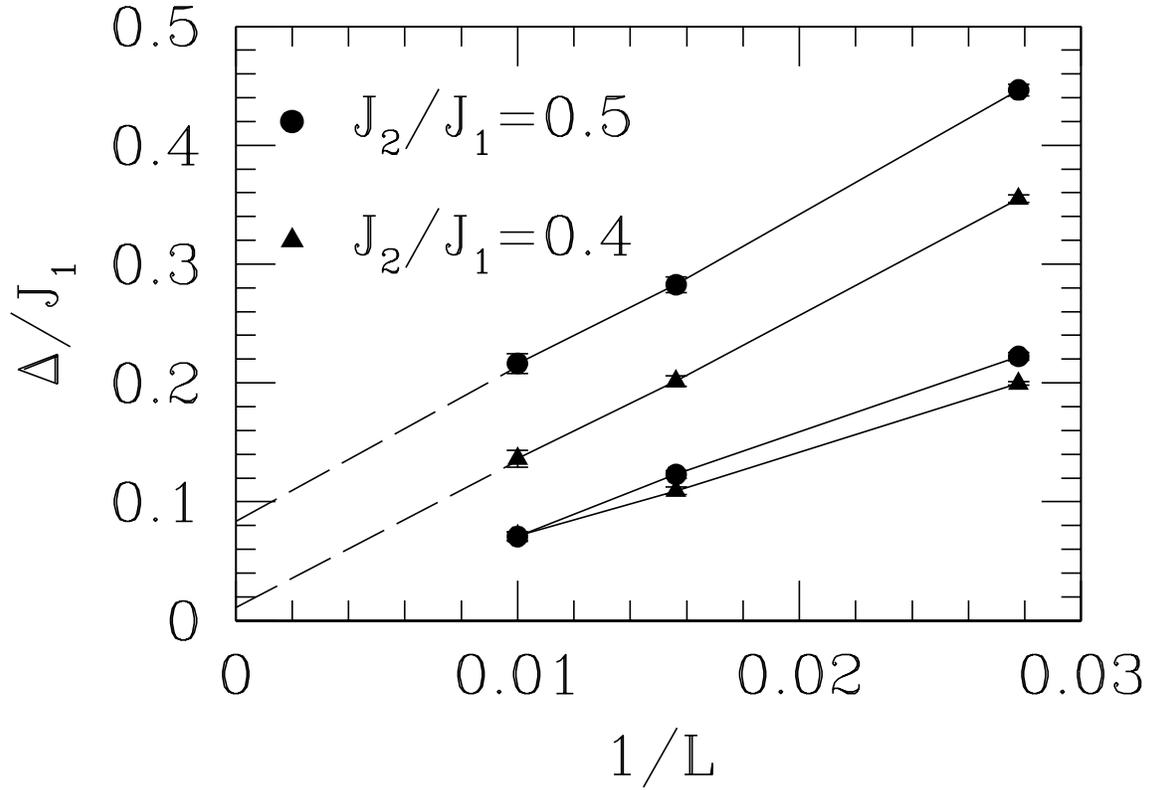,height=11cm}}
\caption {\baselineskip .185in\label{sizee}
   Size scaling of  the spin gap.  
 The dashed lines are  linear  fit
of the GFMCSR data with  $p=9,14,20$ for $L=36,64,100$ respectively. 
Lower curves are the variational estimates and 
continuous lines are guides to the eye.}
\end{figure}

 The calculation was therefore  extended to the large size system  up to  
 $L=100$
 where exact diagonalization is not possible. The spin gap 
as  a function of the system size is displayed in Fig.(\ref{sizee}). 
This figure is 
consistent  with  the opening of a finite 
  spin gap for $J_2/J_1 \ge\sim 0.4$. This gap  is certainly not  an 
artifact of the variational WF, which is obviously gapless, 
as also confirmed numerically in the same figure. 
 The present numerical results confirm 
 that the transition to a spin liquid state
with a finite spin gap but no classical order parameter 
should be close to $J_2/J_1=0.4$.\cite{poilblanc}

  This work was supported in part by INFM (PRA HTSC) and 
CINECA  grant.

  \begin{table}[b]  
  \begin{tabular}{|c|c|c|c|c|c|c|} \hline\hline 
 $J_2/J_1$ & L &  $\eta$ & \% VMC  &  \%  FN  &\% SRe 
 & \% SR \\ \hline 
         0.1 & 16 & 1.2 &     2.84 (2.2) &    0.17 (0.1) &  -0.03 (0.0)  & 0.02
(0.0) \\ \hline 
         0.2  & 16 & 1.15 &    2.80 (2.5)&    0.41 (0.4)     & 0.00 (0.2)    &
0.03 (0.0) \\ \hline 
         0.3  & 16 &  1.1 &  3.25 (2.5)  &   0.87 (0.7)      &   0.12 (0.8) & 
0.05 (0.1)\\ \hline 
         0.4 & 16 &  0.8 &  3.38 (2.4) &    1.76 (3.2)  &   0.56 (4.5) &   0.26
(0.2) \\ \hline 
         0.5   & 16 & 0.85 & 5.65 (10.9)  &    3.84 (8.9)  &    2.08 (8.9)     
&  0.66 (1.1)\\ \hline
         0.1   & 32 & 1 &1.55 (2.5) &    0.22 (0.3) &   0.05 (0.1) &    0.02
(0.0)  \\ \hline  
         0.2   & 32 & 1 & 1.78 (2.5)   &    0.48 (0.6)  &   0.15 (0.6)&    0.05 
(0.1) \\ \hline 
         0.3   & 32 &  1&  2.23 (2.1) &  0.85 (0.91) &      0.30  (1.4)  &  
0.10 (0.0)  \\ \hline 
         0.4   & 32 &  0.8 & 3.07 (4.0) &     1.61 (3.1)  &  0.26 (5.6) &   
0.21 (0.1) \\ \hline 
         0.5   & 32 & 0.9 &   4.51 (10.0) &    2.92 (7.2) &    1.52 (7.7) &   
0.46 (0.9) \\ \hline 

         0.1   & 36 & 1.1 & 1.86 (2.8) &   0.21 (0.2) &   0.1 (0.12) &   0.02
(0.1) \\ \hline 
         0.2   & 36 & 1.1 & 2.22 (2.8)  &  0.47 (0.5) &    0.16  (0.5) &  
0.07 (0.1) \\ \hline 
         0.3   &36 & 1 &  2.31 (2.8)  &  0.91 (1.4) &   0.35 (2.0) &   0.11
(0.1)  \\ \hline 
         0.4   &36 & 0.8 & 3.34 (5.5) & 1.74 (4.5) &  0.51 (6.8) &  0.26 (0.3) 
\\ \hline 
         0.5   &36 & 0.9 &  5.09 (14.4)  & 3.34 (11.1)&  1.83 (11.8) & 0.62
(2.1)  \\ \hline 
  \end{tabular}
  \caption[...]{ 
Percentage error of the energy (square antiferromagnetic order parameter $\vec
m^2$ as in \protect\cite{calandra}) 
  for  the various methods: variational (VMC), fixed node (FN) 
,   $p=1$ GFMCSR  (SRe)   with 
the energy alone and    $p=5,8,9$ GFMCSR estimate (SR) with the 
energy and $S^z_q$  
 for $L=16,32,36$. The statistical errors are 
about   one place in the last digit.}
\label{taben} \end{table}

\begin{references}
\bibitem{theorem} It is possible to prove that the method is variational 
also for $\gamma>0$, with a simple extension of the proof in  \cite{ceperley1}.
\bibitem{guiding} All the analysis remains unchanged if a guiding 
function $\psi_G (x)$ is used for importance sampling. The 
matrix elements of all the operators  $O_{x^\prime,x}$ 
(including the GF)  have to be accordingly changed 
: $ O_{x^\prime,x} \to  \psi_G(x^\prime)  O_{x^\prime,x } / \psi_G(x) $.
\end{references}
\end{document}